\documentclass[conference]{IEEEtran}
\IEEEoverridecommandlockouts
\usepackage{cite}
\usepackage{amsmath,amssymb,amsfonts}
\usepackage{algorithmic}
\usepackage{graphicx}
\usepackage{textcomp}
\usepackage{subcaption}
\usepackage{xcolor}
\definecolor{amber}{rgb}{1.0, 0.75, 0.0}
\definecolor{almond}{rgb}{0.94, 0.87, 0.8}
\definecolor{blond}{rgb}{0.98, 0.94, 0.75}
\definecolor{cornflowerblue}{rgb}{0.39, 0.58, 0.93}
\definecolor{lavenderblue}{rgb}{0.8, 0.8, 1.0}
\definecolor{lightskyblue}{rgb}{0.53, 0.81, 0.98}
\definecolor{lime(web)(x11green)}{rgb}{0.0, 1.0, 0.0}
\definecolor{lime(colorwheel)}{rgb}{0.75, 1.0, 0.0}
\definecolor{persianpink}{rgb}{0.97, 0.5, 0.75}
\definecolor{mistyrose}{rgb}{1.0, 0.89, 0.88}
\definecolor{ticklemepink}{rgb}{0.99, 0.54, 0.67}
\definecolor{salmonpink}{rgb}{1.0, 0.57, 0.64}
\definecolor{richbrilliantlavender}{rgb}{0.95, 0.65, 1.0}
\definecolor{pink}{rgb}{1.0, 0.75, 0.8}
\definecolor{cadetgrey}{rgb}{0.57, 0.64, 0.69}
\definecolor{darkpastelblue}{rgb}{0.47, 0.62, 0.8}
\DeclareMathOperator*{\argmax}{arg\,max}

\def\BibTeX{{\rm B\kern-.05em{\sc i\kern-.025em b}\kern-.08em
    T\kern-.1667em\lower.7ex\hbox{E}\kern-.125emX}}
\begin{document}

\title{Measurement-Based Modeling of Short Range Terahertz Channels and Their Capacity Analysis}

\author{\IEEEauthorblockN{Onur Salan\IEEEauthorrefmark{1}, 
Ferhat Bayar\IEEEauthorrefmark{2}\IEEEauthorrefmark{3},
Hacı Ilhan\IEEEauthorrefmark{3}, 
Erdogan Aydin\IEEEauthorrefmark{4}}

\IEEEauthorblockA{\IEEEauthorrefmark{1} Communications and Signal Processing Research (HISAR) Laboratory, T{U}B{I}TAK B{I}LGEM, Kocaeli, Turkey}

\IEEEauthorblockA{\IEEEauthorrefmark{2} Scientific and Technological Research Council of Turkey T{U}B{I}TAK B{{I}}LGEM, Kocaeli, Turkey} 

\IEEEauthorblockA{\IEEEauthorrefmark{3} Department of Electronics and Communications Engineering, Yildiz Technical University, {I}stanbul, Turkey} 

\IEEEauthorblockA{\IEEEauthorrefmark{4} Department of Electrical and Electronics Engineering, Istanbul Medeniyet University, {I}stanbul, Turkey}
\texttt{onur.salan@tubitak.gov.tr, ferhat.bayar@tubitak.gov.tr} \\
\texttt{ilhanh@yildiz.edu.tr, erdogan.aydin@medeniyet.edu.tr}}

\title{Deep Learning Based Detection on RIS Assisted RSM and RSSK Techniques}
\maketitle

\begin{abstract}
The reconfigurable intelligent surface (RIS) is considered a crucial technology for the future of wireless communication. Recently, there has been significant interest in combining RIS with spatial modulation (SM) or space shift keying (SSK) to achieve a balance between spectral and energy efficiency. In this paper, we have investigated the use of deep learning techniques for detection in RIS-aided received SM (RSM)/received-SSK (RSSK) systems over Weibull fading channels, specifically by extending the RIS-aided SM/SSK system to a specific case of the conventional SM system. By employing the concept of neural networks, the study focuses on model-driven deep learning detection namely block deep neural networks (B-DNN)  for RIS-aided SM systems and compares its performance against maximum likelihood (ML) and greedy detectors. Finally, it has been demonstrated by Monte Carlo simulation that while B-DNN achieved a bit error rate (BER) performance close to that of ML, it gave better results than the Greedy detector.
\end{abstract}

\section{Introduction}
Future wireless technologies, including the fifth-generation technology standard (5G) and beyond, aim to achieve much higher data rates, energy-efficient protocols, and hardware simplicity due to the significant growth in user numbers, applications, and services\cite{andrews2014will}. In this context, there is a great interest in index modulation (IM) techniques, which are digital modulation techniques that offer high spectral and energy efficiency with low complexity. Unlike conventional communication systems that transmit information using the amplitude, frequency, or phase of the signal, IM techniques utilize the activity information of transmission elements, such as modulation type, signal power, and antenna information, for efficient information transmission \cite{basar2017index}. Spatial modulation (SM) and space shift keying (SSK) have also been actively explored in recent literature. SM, a modulation technique that utilizes the spatial domain, has gained attention due to its ability to increase the data transmission rate and improve spectral efficiency. Researchers have investigated the application of SM in various wireless communication scenarios, including massive multiple-input multiple-output (MIMO) systems, visible light communication, and millimeter-wave communication. Likewise, SSK which relies on intelligent phase shifting, has shown promise in enhancing communication performance by leveraging multiple antennas.

One of the assumptions in classical communication theory is that events occurring in wireless environments are random processes. At this point, the concept of a smart radio environment (SRE) has been proposed, where the reflection of signals can be controlled. In a smart radio environment, the power of the signal reflected towards the user is maximized, while the power of the signal reflected in other directions is minimized. RISs have emerged as promising technologies in the field of wireless communications. In this context, RISs are proposed as intelligent devices that dynamically control the propagation environment to enhance the signal quality at the receiver \cite{liaskos2018new, di2019smart, basar2019wireless, danufane2021wireless}. RISs are surfaces composed of a large number of passive reflecting elements that can be dynamically adjusted to modify the characteristics of the wireless channel. By intelligently manipulating the reflections, RISs can improve signal strength, mitigate interference, and enhance the overall performance of wireless communication systems. 

Researchers have focused on investigating the benefits of integrating RIS with existing modulation schemes, such as SM and SSK \cite{basar2020reconfigurable, salan2021performance, canbilen2020reconfigurable}. These studies have highlighted the potential of RIS to enhance both spectral and energy efficiency by intelligently manipulating the wireless propagation environment. The analysis of recent literature suggests that RIS, when combined with SM or SSK, can offer improved performance and provide a trade-off between spectral efficiency and energy consumption in wireless communication systems.

Block deep neural networks are utilized in the SM technique to reduce the active antenna indices and symbol estimation processing time in \cite{albinsaid2020block}. Furthermore, recent literature analysis reveals a growing interest in the application of deep learning for RIS-aided SM/SSK systems, focusing on model-driven and data-driven approaches \cite{liu2021data}. Researchers have leveraged the power of deep neural networks to develop advanced detection algorithms that can effectively handle the complex nature of RIS-aided communication systems\cite{10025789}.

In this study, we employed the B-DNN algorithm to reduce processing complexity at the receiver in the RIS-based SM/SSK system model over the Weibull fading channel. The Weibull fading channels provide a more precise representation of the variability and diversity of the wireless channel, offering realistic results particularly in the performance analysis of advanced technologies such as RISs. Subsequently, we compared the error performance of the receiver B-DNN using the obtained results with the receiver structures of ML and greedy algorithms. Additionally, the choice of utilizing the B-DNN algorithm in this study is driven by its capability to alleviate the high complexity in wireless communication and expedite symbol estimation processes, aiming to enhance system performance. Parameters/symbols used in this paper are represented in Table \ref{TableListOfSymbols}. 

\subsection{Paper Organization}
The remainder of this paper is organized as follows. Section II outlines the RIS-based Received SM/SSK schemes with a DNN-based receiver and defines the receiver structure. The Monte Carlo simulation results are presented in Section IV. Finally, Section V concludes the paper.

\begin{table}
  \caption{List of Parameters/Symbols}
\centering
\scalebox{0.99}{%
\begin{tabular}{|l |c |c|}
		
\hline\hline 
Parameters/Symbols  & Definition\\\hline\hline 
$\mathrm{E}\left[X\right](\mu_X)$  & Mean of RV \textit{X}
\\ 
\hline

$\mathrm{Var}\left[X\right](\sigma^2_{X})$  & Variance of RV \textit{X} 
\\ 

\hline

$f_X\left(\cdot\right)$ & Probability density function (PDF) of RV \textit{X}
\\
\hline

$\exp\left(\cdot\right)$ & Exponential function
\\
\hline

$N_{0}$ & Variance of AWGN noise
\\
\hline

$\gamma$ & Instantaneous signal to noise ratio (SNR)
\\
\hline
\end{tabular}}
\label{TableListOfSymbols}

\end{table}

\section{System Model}
\begin{figure*}
\centering
\includegraphics[width=\textwidth]{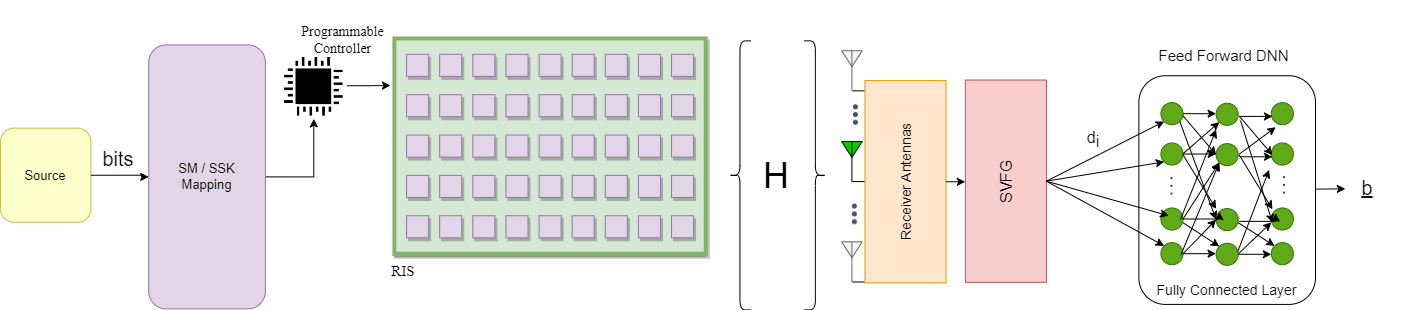}
\caption{System model of RIS-based SM/SSK using B-DNN at receiver}
\label{fig:GeneralSystemModel}
\end{figure*}
We employ the idea of using the RIS itself as an access point (AP), as described in \cite{salan2021performance}. In this concept, the signals emitted by a nearby RF source bounce off a wall or surface that consists of $N$ passive reflector elements. These signals are then directed towards the destination, which is equipped with $N_r$ receiving antennas. It is important to note that the destination is located far away from the RIS and does not directly receive signals from the RF source. The RIS adjusts the phase $\Phi_{\iota}$ of each reflector element based on the channel phase, effectively reflecting and improving the incoming signal from the RF source to enhance error performance. In this model, we assume that channel state information between RIS and destination is known at the RIS controller. In order to increase instantaneous signal-to-noise ratio (SNR) at the $m^{th}$ receive antenna, $\Phi_{\iota}=\theta_{m,\iota}$ for $\iota = 1,2,\dots,N$ is adjusted according to the information bits generated from the sources. Therefore, we derive instantaneous SNR at the $m^{th}$ selected receive antenna as follows:
\begin{equation}
    \gamma_{m}= \frac{\left\vert\sum_{{\iota} = 1}^{N}\beta_{m,{\iota}} e^{j(\Phi_{{\iota}}-\theta_{m,{\iota}})} \right\vert^{2}E_s}{N_0} = \frac{\left\vert\sum_{{\iota} = 1}^{N}\beta_{m,{\iota}}\right\vert^{2}E_s}{N_0}
\end{equation}
where $E_s$ is the energy per symbol and $N_0$ is the variance of noise.
The wireless fading channel between $l$th receive antenna and ${\iota}$th reflector element is characterized by
\begin{equation}
g_{l,{\iota}}=\beta_{l,{\iota}}e^{-j\theta_{l,{\iota}}} \quad l= 1, 2, \cdots, \textit N_R, \quad {\iota}= 1, 2, \cdots, \textit N
\label{CharacterizedChannelFormula}
\end{equation}
where $\theta_{l,{\iota}}$ is the channel phase induced by the ${\iota}$th reflector at the $l$th receive antenna and $\beta_{l,{\iota}}$ is channel fading coefficients between the ${\iota}$th reflector and $l$th receive antenna follow the Weibull distribution. 

Furthermore, we performed symbol and antenna estimation by utilizing deep neural network at the receiver. For enhanced the performance of the symbol classification process, we employed the concept of a separate feature vector generator (SFVG) defined in \cite{albinsaid2020block} during the data pre-processing stage. This approach allowed us to convert the complex-valued IQ raw data into a clean dataset, leading to improved speed and accuracy.

\subsection{ML Detection}
The ML detector, which is an optimal detector, employs the following formula to estimate the transmitted symbols $x$ and antenna index \cite{basar2020reconfigurable}:
\begin{equation}
    \Big(\hat{x},\hat{m}\Big) =  \arg \min\limits_{x,m}\sum\limits_{l=1}^{N_r}{\Big\vert\Big\vert {y}_{l} - \left(\sum\limits_{{\iota}=1}^{N}{\beta_{l,{\iota}}}e^{j(\theta_{m,{\iota}}-\theta_{l,{\iota}})}\right){x}\Big\vert\Big\vert^2}, 
    \label{Estimated Signal}
\end{equation}
where $\hat{x}$ represents the estimated transmitted symbols, and $\hat{m}$ denotes the estimated antenna index and $y_l$ refers to received signal at the $lth$ antenna. 

\subsection{Greedy Detection}
At the receiver, the Greedy detector estimates the receive antenna index as the maximum instantaneous SNR on the receiver without channel estimation:
\begin{equation}
    \hat{m} = \argmax\limits_{m}|{y_l}|^2,
    \label{Estimated Signal at RIS/SSK}
\end{equation}
whereby $m$ and $\hat{m}$ are the selected and estimated receive antenna index at the destination (D), respectively. After the detection of the receive antenna index, the transmitted data symbol is detected and estimated as follows:
\begin{equation}
    \hat{x} = \arg \min\limits_{x}{\left\vert y_{\hat{m}} - \left(\sum\limits_{{\iota}=1}^{N}{\beta_{\hat{m},{\iota}}}\right)x\right\vert^2}.
    \label{Estimated Signal at RIS/SM}
\end{equation}
\subsection{Block DNN Detection}
The crucial factor in improving the effectiveness of a deep learning algorithm lies in data pre-processing. Data pre-processing involves applying various modifications to the input data before feeding it into the DNN model. Here is a future extraction method:
\begin{equation}
\begin{aligned}
f_{SFVG} \left(y\right) &=
\\ &\left[\left|\Re(y_{1,1})\right|,\left|\Im(y_{1,1})\right|,\cdots,\left|\Re(y_{N_r,1})\right|,\left|\Im(y_{Nr,1})\right|\right]^T
\end{aligned}
\end{equation}

The feature extraction vector is formed by concatenating the received signal and the channel matrix. Since both the received signal and the channel matrix are expressed in complex form, we sequentially convert their real and imaginary components into a vector array using the following procedure:
\begin{equation}
d_i= \left[f_{SFVG}\left(y^{j}\right)^T, f_{SFVG}\left(H^{j}\right)^T\right]^T.
\end{equation}

\begin{table}[!ht]
\centering
\caption{Network Parameters and Traning Configuration }
\begin{tabular}{|c|c|}
\hline
\textbf{Parameters} & \textbf{Values}  \\
\hline
Input Nodes & 2($N_r$+$N_r$)
\\
\hline
Learning Rate & 0.005 \\
\hline
Hidden Layer Activation & ReLU 
\\
\hline
Output Layer Activation & Softmax 
\\
\hline
Loss Function & Cross-Entropy 
\\
\hline
Optimization & SGD  
\\ 
\hline
Epoch & 50 
\\
\hline
BPSK Hidden Nodes & 128-64-32  
\\ 
\hline
QPSK Hidden Nodes & 256-128-64
\\
\hline QAM Hidden Nodes & 512-256-128  \\
\hline
\end{tabular}
\label{table:DNNParameters}
\end{table}

For the estimation of each received signal, the proposed DNN utilizes $L$ fully connected layers with $L - 1$ hidden layers. The parameters of the recommended DNN are presented in Table \ref{table:DNNParameters}. The input and output mappings of the $L$-layered DNN series are expressed as follows
\begin{equation}
\begin{aligned}
Z_L =&\sigma\left(W^{L}\left(\sigma\left(W^{L-1}\left(\cdots\sigma\left(W^{1}Z_{0}+b_{1}\right)\cdots\right)+b_{L-1}\right)\right)\right.
\\
&\left.+b_L\right)
\end{aligned}
\end{equation}
Here, $\sigma$ represents the activation function, $W^{\left(l\right)}$ is the weight matrix, $b_l$ denotes the bias vector, and $Z_0$ corresponds to the final vector input, which is equal to $d_i$. The Rectified Linear Unit(ReLU) function, which is employed in the layers excluding the last layer, can be defined as $\sigma(x) = \max(0, x)$. The activation function outputs zero when the input value is negative and directly outputs the input value when it is positive. The softmax function, used in the last layer, assigns probability values to the neurons in the range of $\left[0, 1\right]$ ensuring that their sum is equal to $1$. 

One commonly used loss function in classification problems is the cross-entropy function. It calculates the difference between the predicted probabilities and the true data to obtain an error score. To minimize the loss function, the stochastic gradient descent (SGD) algorithm is utilized. SGD optimizes the loss function by iteratively updating the model parameters. The training process continues for a certain number of iterations or until another stopping criterion is met. During each iteration, SGD computes the gradients of the loss with respect to the parameters and updates the parameters in the direction that reduces the loss. This iterative process helps the model converge towards the optimal set of parameters for minimizing the loss and improving the overall performance of the model.

\begin{figure}
\centering
\includegraphics[width=0.5\textwidth]{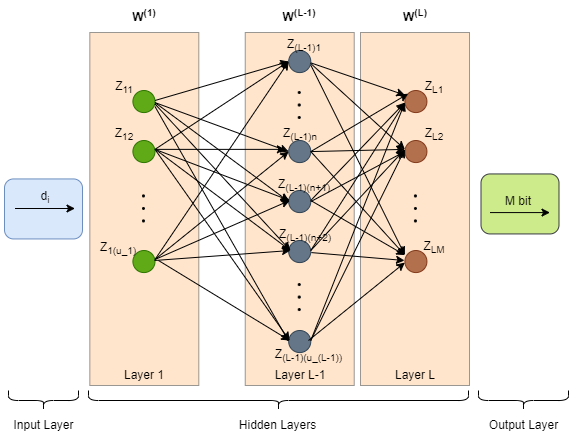}
\caption{Fully connected layers}
\label{fig:BDNNSystemModel}
\end{figure}
During the estimation phase, for each transmission time, the $N$ input vectors obtained, denoted as $d_i$, generate the output vector $Z_L$. In this context, the transmitted symbol is estimated as $\hat{x}$ = $S_n$, using the following equation:

\begin{equation}
\hat{n} = \underset{{n \in {1,\cdots,M}}}{\arg\max}\left(Z_{L_n}\right)
\end{equation}
The estimation results in $K$ symbol vectors, and the index of the minimum value is determined by calculating the Euclidean distance between each of these symbol vectors and the received signal at the receiver. The index corresponding to the minimum distance represents the estimated symbol:
 \begin{equation}
        \hat{i} = \underset{{i \in {1,\cdots,K}}}{\arg\min}\lVert \mathbf{y}^j - \mathbf{H}_{l}^j\hat{x}^j \rVert_{F}^2
    \end{equation}
Lastly, the symbol expression corresponding to the obtained index value becomes the estimated output symbol of the B-DNN receiver structure.

\section{Simulation Results}
\begin{figure}
     \centering
\includegraphics[width=0.5\textwidth]{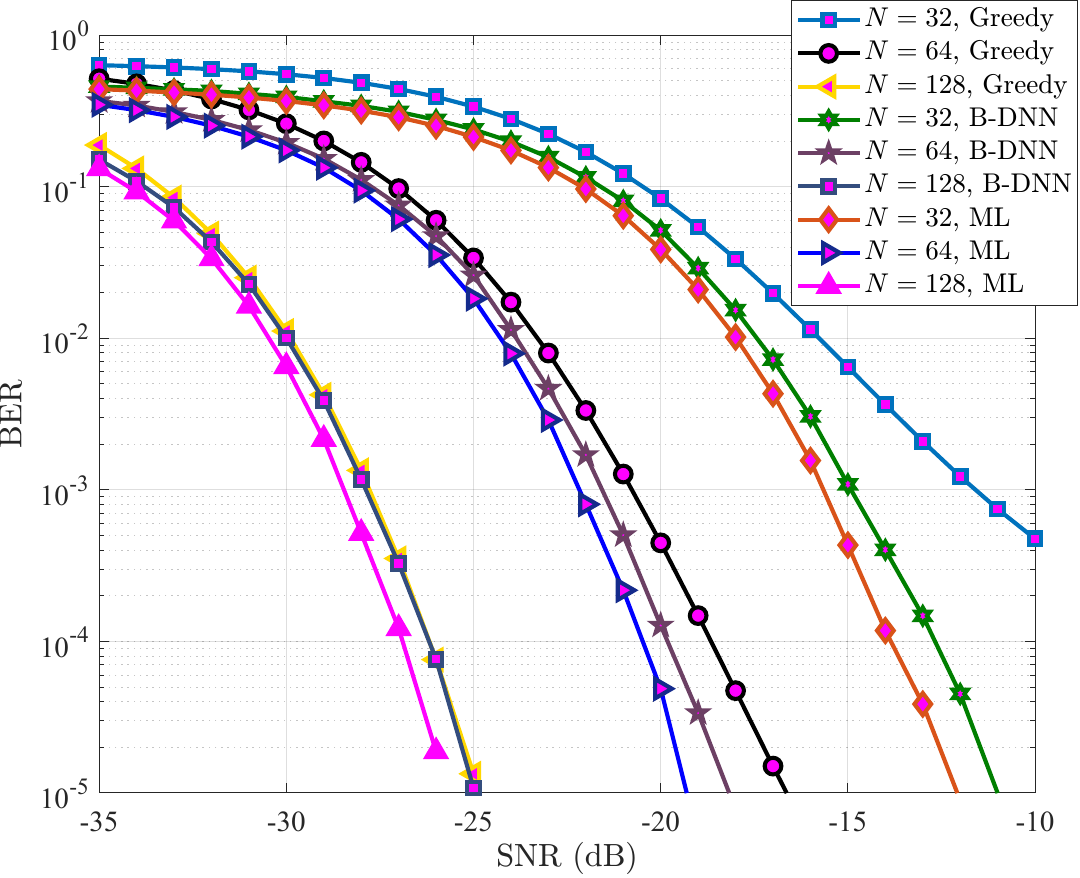}
     \caption{BER performance comparison of RIS-SM between ML, Greedy and B-DNN receivers for varying $N$ ($N_r = 4$, $M = 4$, $\alpha = 1.2$, $\Omega = 1$)}
     \label{Fig:Varying N for SM}
\end{figure}

This section verifies the analytical error performance of the proposed system model using Monte Carlo simulation results. All simulation results are obtained under the assumption of independent Weibull fading channels with severity of the fading $\alpha = 1.2$ and average fading power $\Omega=1$. As depicted in Fig.\ref{Fig:Varying N for SM}, the system's error performance demonstrates improvement with an increasing number, $N$, when the receiver adopts the B-DNN. Notably, the B-DNN receiver significantly enhances the system's error performance, particularly for $N = 32$, approaching the error floor observed in the greedy receiver. Additionally, the impact of the SNR increase on performance gradually diminishes when employing a lower number of reflective surfaces in the greedy receiver. However, the utilization of the B-DNN receiver eliminates the error performance limitations associated with the number of surfaces. B-DNN detection has exhibited similar BER performance to the greedy approach and approaches the performance of the ML detector at $N=128$.

\begin{figure}
     \centering
     \includegraphics[width=0.5\textwidth]{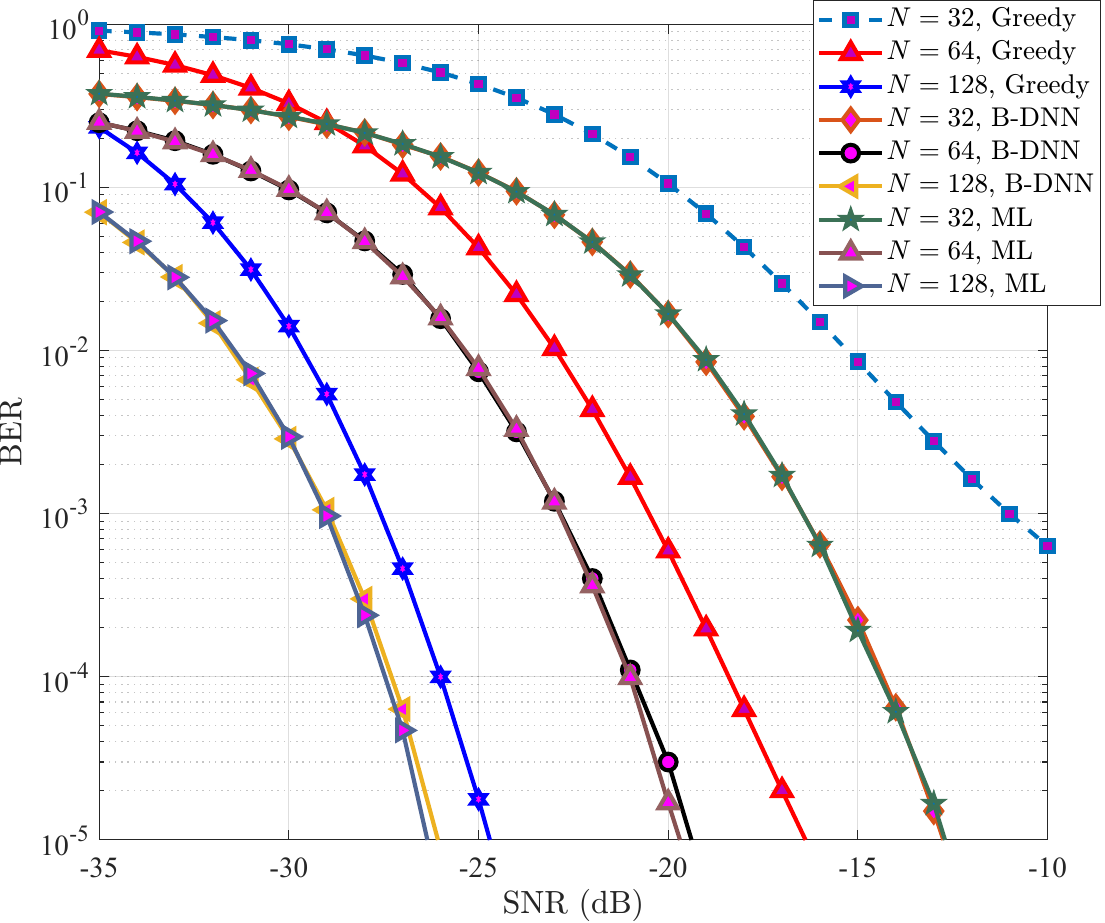}
     \caption{BER performance comparison of RIS-SSK between ML, Greedy and B-DNN receivers for varying $N$ ($N_r = 4$, $\alpha = 1.2$, $\Omega = 1$).}\label{Fig:Varying N for SSK}
\end{figure}

In Fig.\ref{Fig:Varying N for SSK}, the positive influence of the B-DNN receiver on the error performance of RIS-SSK is observed more prominently. Furthermore, similar to the greedy receiver, the increasing number of reflective surfaces also exhibits a beneficial impact on the bit error rate performance in the B-DNN receiver. It is clearly seen that B-DNN detection outcomes nearly same BER performance with ML detection for each $N$ cases.

\begin{figure}
     \centering
     \includegraphics[width=0.5\textwidth]{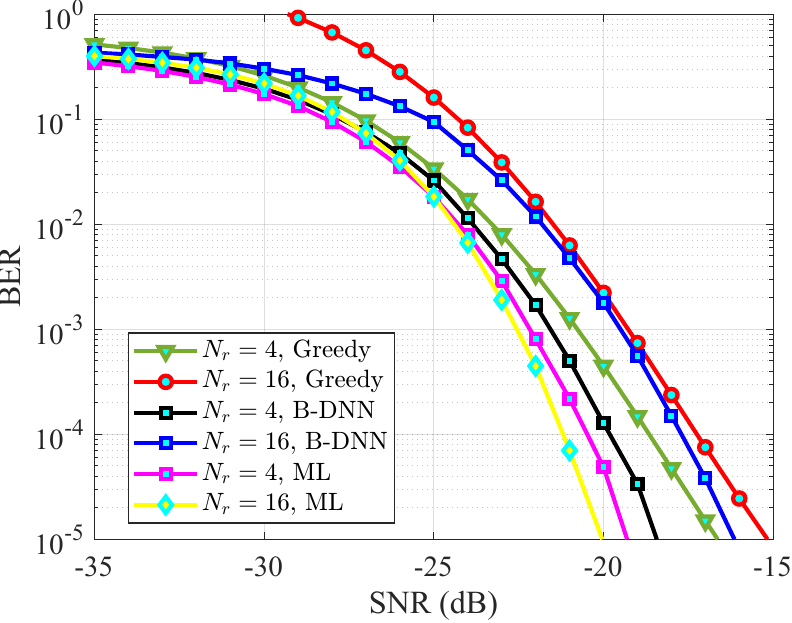}
     \caption{BER performance comparison of RIS-SM between ML, Greedy and B-DNN receivers for varying $N_r$ ($N = 64$, $\alpha = 1.2$, $\Omega = 1$).}
     \label{Fig:Varying Nr for SM}
\end{figure}

Fig.\ref{Fig:Varying Nr for SM} presents the impact of the number of receiver antennas on RIS/SM with B-DNN, Greedy, and ML.
In this system configuration, B-DNN showed similar result to Greedy. In other words, the system performance deteriorated with increasing Nr for B-DNN and greedy. However, B-DNN still has better BER performance than greedy.

\begin{figure}
     \centering
     \includegraphics[width=0.5\textwidth]{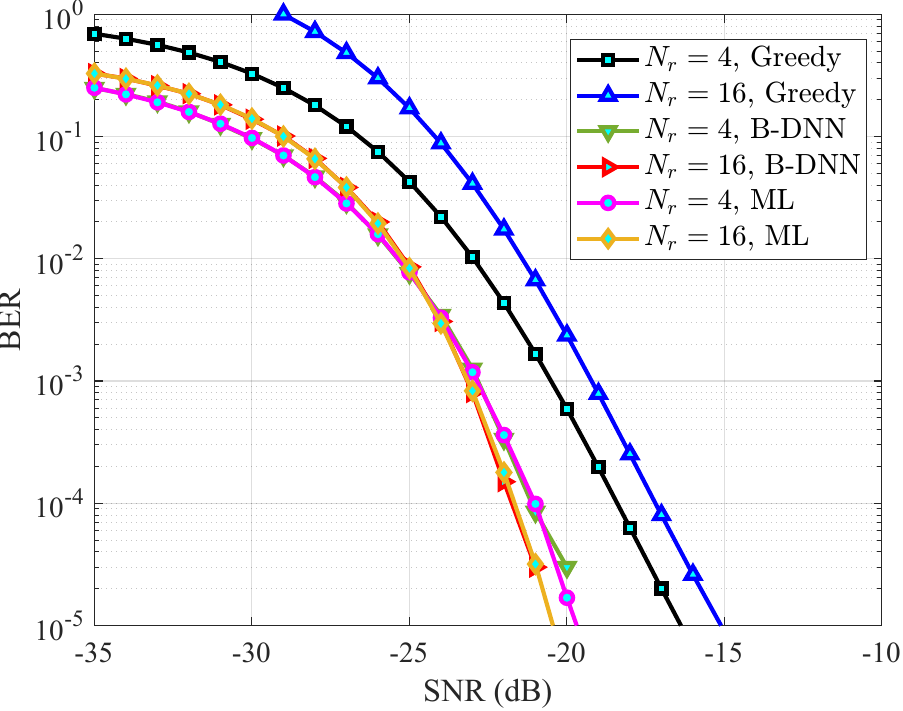}
     \caption{BER performance comparison of RIS-SSK between ML, Greedy and B-DNN receivers for varying $N_r$ ($N = 64$, $\alpha = 1.2$, $\Omega = 1$).}
     \label{Fig:Varying Nr for SSK}
\end{figure}

Fig.\ref{Fig:Varying Nr for SSK} illustrates the effect of number of receiver antenna on the error performance of RIS/RSKK for each detectors. It is clearly observed that as increase in $N_r$ improves the BER performance\cite{salan2021performance,basar2020reconfigurable} for both ML. It is can also seen that B-DNN performs same behaviour as ML. Plus, B-DNN gives better BER results than greedy. 

\section{Conclusion and Future Works}
In future studies, the importance of autonomous and self-paced systems is increasing day by day. Consequently, artificial intelligence-based technologies are being frequently employed in communication systems. In this paper, we employed the B-DNN structure in the receiver to enable autonomous communication. We estimated the symbol transmitted in Weibull channels and the antenna indexes carrying additional information bits using the B-DNN. Additionally, we improved data rates using SM and SSK technologies. In the proposed model, we consider the RIS as an access point situated between the source and destination, where the fading follows a Weibull distribution. We compare various receivers, including ML, Greedy, and a deep learning-based B-DNN. Simulation results demonstrate that the BER performance of the B-DNN detector closely matches that of the ML receiver, especially in RIS-based SSK schemes. Additionally, the B-DNN outperforms the greedy receiver in both schemes.

\vspace{12pt}

\bibliographystyle{IEEEtran}
\bibliography{IEEEabrv,referanslar}
\end{document}